\pdfoutput=1
\documentclass[conference]{IEEEtran}
\IEEEoverridecommandlockouts
\usepackage{cite}
\usepackage{amsmath,amssymb,amsfonts}
\usepackage{algorithmic}
\usepackage{graphicx}
\usepackage[T1]{fontenc}% optional
\usepackage{textcomp}
\usepackage{xcolor}
\usepackage{bm}

\def\BibTeX{{\rm B\kern-.05em{\sc i\kern-.025em b}\kern-.08em
    T\kern-.1667em\lower.7ex\hbox{E}\kern-.125emX}}

\begin{document}

\title{ISAC-Enabled V2I Networks Based on 5G NR: How Much Can the Overhead Be Reduced?}

\author{
\IEEEauthorblockN{
Yunxin Li\IEEEauthorrefmark{1}, 
Fan Liu\IEEEauthorrefmark{1}, 
Zhen Du\IEEEauthorrefmark{1}\IEEEauthorrefmark{2}, 
Weijie Yuan\IEEEauthorrefmark{1}, 
Christos Masouros\IEEEauthorrefmark{3}}
\IEEEauthorblockA{\IEEEauthorrefmark{1}Southern University of Science and Technology, Shenzhen, China}
\IEEEauthorblockA{\IEEEauthorrefmark{2}Nanjing University of Information Science and Technology, Nanjing, China}
\IEEEauthorblockA{\IEEEauthorrefmark{3}University College London, London, UK}
\IEEEauthorblockA{liyx2022@mail.sustech.edu.cn, \{liuf6, yuanwj\}@sustech.edu.cn, duzhen@nuist.edu.cn, c.masouros@ucl.ac.uk}}
   
\maketitle

\begin{abstract}
The emergence of the fifth-generation (5G) New Radio (NR) brings additional possibilities to vehicle-to-everything (V2X) network with improved quality of services. In order to obtain accurate channel state information (CSI) in high-mobility V2X networks, pilot signals and frequent handover between vehicles and infrastructures are required to establish and maintain the communication link, which increases the overheads and reduces the communication throughput. To address this issue, integrated sensing and communications (ISAC) was employed at the base station (BS) in the vehicle-to-infrastructure (V2I) network to reduce a certain amount of overheads, thus improve the spectral efficiency. Nevertheless, the exact amount of overheads reduction remains unclear, particularly for practical NR based V2X networks. In this paper, we study a link-level NR based V2I system employing ISAC signaling to facilitate the communication beam management, where the Extended Kalman filtering (EKF) algorithm is performed to realize the functions of tracking and predicting the motion of the vehicle. We provide detailed analysis on the overheads reduction with the aid of ISAC, and show that up to 43.24\% overheads can be reduced under assigned NR frame structure. In addition, numerical results are provided to validate the improved performance on the beam tracking and communication throughput.
\end{abstract}


\section{Introduction}
V2X networks are envisioned as a key enabler for future intelligent transportation applications, which largely rely upon the seamless ultra-low latency wireless connections provided by the next-generation cellular system, namely, 5G-Advanced (5G-A) and 6G\cite{wymeersch20175g}. In addition to the communication capability, the future V2X networks are also anticipated to have highly reliable and accurate active sensing functionality to support a variety of environment-aware services, such as simultaneous localization and mapping (SLAM) and vehicle platooning. At the time of writing, the localization and positioning services for V2X networks are mainly built on the global navigation satellite-based systems (GNSS), which however suffer from poor resolution and low refresh rate, and thus fail to fulfill the demanding requirement of the future vehicular applications\cite{wymeersch20175g}. More severely, the high mobility of the V2X network imposes critical challenges in channel training and beamforming, which requires frequent coordination and feedback between the BS/road side unit (RSU) and the vehicle. This inevitably generates considerable signaling overheads, and thereby results in performance loss in the communication throughput. 

To tackle the above issues, ISAC, which simultaneously implements active sensing and communications functionalities through exploiting both the massive multi-input-multi-output (mMIMO) antenna array and millimeter wave (mmWave) technologies of the 5G NR infrastructures, is well-recognized as a promising solution\cite{liu2020joint}. In particular, finer angular and range resolution can be achieved by scaling up the antenna array and increasing the bandwidth, respectively. More remarkably, it was reported that the ISAC signaling is capable of significantly reducing the channel estimation overheads as compared to communication-only mmWave beam training and tracking, since it requires no dedicated downlink pilots, and removes the uplink feedback via directly processing the reflected echoes from the vehicles. This technique, which is referred to as \textit{sensing-assisted communications}, has recently gained tremendous attention from the research community. 

Most of the existing works on ISAC-enabled V2X networks mainly concentrate on the algorithmic design aspect of sensing-assisted beam training and tracking. Pioneered by \cite{liu2020radar,yuan2020bayesian}, ISAC based V2X beamforming design was proposed by leveraging various Bayesian filtering approaches, such as the EKF and message passing algorithm. To track the extended vehicular objects, an ISAC-based alternating beamforming (ISAC-AB) technique was conceived by dynamically changing the beamwidth at the BS according to the distance of the vehicles \cite{du2022integrated}. More recently, a curvilinear coordinate system (CCS) based ISAC beamforming method was designed to serve vehicles driving on complex roads \cite{meng2022vehicular}, e.g., roundabout and highway overpass. While these schemes were well-designed with sophisticatedly tailored algorithms, they generally assume simplistic frame structures and transmission protocols, and are thus difficult to be straightforwardly applied to the 5G NR framework. More importantly, the exact amount of overheads reduction that can be achieved in practical systems still remains unclear.

To fill in this research gap, in this paper we investigate a link-level 5G NR based vehicle-to-infrastructure (V2I) system employing ISAC signaling to reduce the overheads. More specifically, we consider an mMIMO ISAC BS/RSU that serves a vehicle in the downlink, by transmitting standardized orthogonal frequency-division multiplexing (OFDM) based ISAC waveforms operating at the 5G NR frequency range two (FR2). We provide a thorough analysis on the NR frame structure and the resource blocks that can be used for realizing the dual functionalities of sensing and communications, and propose an ISAC signal processing framework that exploits the NR signals for vehicle parameter estimation and tracking. The simulation environment is constructed by using ray-tracing method in a real scene in Shenzhen, China, by taking into account both line-of-sight (LoS) and non-LoS (NLoS) paths. Our results convincingly demonstrate the superiority of ISAC based scheme over its communication-only counterpart in terms of the angle tracking accuracy and achievable throughput, and show that up to 43.24\% communication overhead reduction can be achieved thanks to the assistance of the active sensing functionality.

\section{Sensing-assisted communications}

\begin{figure}[htbp]
\centering
\includegraphics[width=\columnwidth]{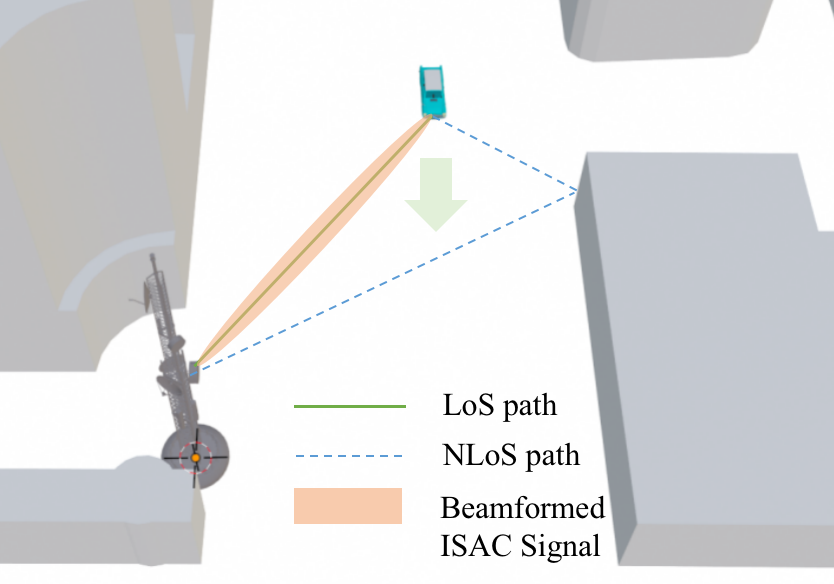}
\caption{Simulation scenario}
\label{fig1}
\end{figure}

Throughout this paper, we consider an mMIMO BS operating in FR2 with uniform planar array (UPA) with half-wavelength spacing, which serves a single vehicle that is also equipped with MIMO UPA. We assume that the vehicle is driving along a straight road, communicating with the BS via both LoS and NLoS channels, as shown in Fig.\ref{fig1}. All the kinematic parameters are defined in the time window of $t \in[0, T_{max}]$, where $T_{max}$ denotes the maximum time duration of interest that can be divided into small time slots $\Delta T$. Accordingly, the angle, distance and velocity of the vehicle at the $n$th time slot are represented as $\bm{\theta}_{n}$, $d_n$ and $v_n$ respectively and are assumed to be constant in each time slot.   

\addtolength{\topmargin}{0.1in}

\subsection{Radar Signal Model}

At the $n$th time slot, the reflected echoes received at BS from the target and K-1 scatterers can be formulated as 
\begin{equation} \label{1}
\begin{aligned}
&\mathbf{r}_n(t)=  \zeta \sqrt{p}\sum_{k=1}^K  \beta_{k, n}  e^{j 2 \pi \mu_{k, n} t} \mathbf{b}(\bm{\theta}_{k, n}) \mathbf{a}^H(\bm{\theta}_{k, n}) \mathbf{f}_{n} \\ 
&\times \sum_{m=0}^{M-1}\sum_{l=0}^{L-1} s_{m,l} e^{j 2 \pi m \Delta f \left(t-lT_{s}-\tau_{k,n} \right)}+\mathbf{z}_r(t)
\end{aligned}
\end{equation}
where $\zeta = \sqrt{N_tN_r}$ denotes the array gain factor, with $N_t$ and $N_r$ being the number of transmit and receive antennas respectively, $p$ denotes the transmitted signal power, and $\beta_{k, n}$ and $\mu_{k, n}$ denote the reflection coefficient and the Doppler frequency of the $k$th scatterer, respectively. The reflection coefficient can be expressed as $\beta_{k, n}=\epsilon_{k, n}(2d_{k, n})^{-2}$ if the complex radar cross-section (RCS) $\epsilon_{k, n}$ and the relative distance $d_{k, n}$ are given. The Doppler frequency $\mu_{k, n}=2v_{k,n}f_{c}c^{-1}$ and the time delay $\tau_{k,n}=2d_{k, n}c^{-1}$ are determined by the radial velocity $v_{k,n}$ and the distance $d_{k,n}$ of the $k$th scatterer respectively. $f_c$, $M$, $L$ and $\Delta f$ denote the carrier frequency, the number of subcarriers, number of symbols and subcarrier spacing in the OFDM signal respectively and $s_{m,l}$ denotes the transmitted data on the $m$th subcarrier of the $l$th symbol. Let $T_{s}=T_{cp}+T$, where $T_{s}$, $T_{cp}$ and $T$ denote the total duration of an OFDM symbol, the cyclic prefix duration and the elementary symbol duration, respectively. Finally, $\mathbf{z}_r(t)$ denotes the complex additive white Gaussian noise with zero mean.

The transmit and receive steering vectors of BS's UPA, namely, $\mathbf{a}\left(\bm{\theta}_{k, n}\right)$ and $\mathbf{b}\left(\bm{\theta}_{k, n}\right)$ in (\ref{1}), can be represented as 
\begin{equation}\label{2}
\mathbf{a}(\bm{\theta}_{k, n})=\mathbf{a}(\theta, \phi)=\mathbf{v}_{az}(\theta, \phi) \otimes \mathbf{v}_{el}(\phi)
\end{equation}
\begin{equation}\label{3}
\mathbf{b}(\bm{\theta}_{k, n})=\mathbf{b}(\theta, \phi)=\mathbf{u}_{az}(\theta, \phi) \otimes \mathbf{u}_{el}(\phi)
\end{equation}
where $\theta$ and $\phi$ are the azimuth and elevation angles, and $\mathbf{v}_{az}(\theta, \phi)$, $\mathbf{v}_{el}(\phi)$ and $\mathbf{u}_{az}(\theta, \phi)$, $\mathbf{u}_{el}(\phi)$ are the transmit and receive steering vectors in the horizontal and vertical directions, respectively.
\begin{equation}
\mathbf{v}_{az}(\theta, \phi)=\sqrt{\frac{1}{N_{t,x}}}\left[1, e^{j \pi \sin \theta \cos \phi},\cdots,e^{j \pi\left(N_{t,x}-1\right) \sin \theta \cos \phi}\right]^T
\end{equation}
\begin{equation}
\mathbf{v}_{el}(\phi)=\sqrt{\frac{1}{N_{t,y}}}\left[1, e^{j \pi \sin \phi}, \cdots, e^{j \pi\left(N_{t,y}-1\right) \sin \phi}\right]^T
\end{equation}
\begin{equation}
\mathbf{u}_{az}(\theta, \phi)=\sqrt{\frac{1}{N_{r,x}}}\left[1, e^{j \pi \sin \theta \cos \phi}, \cdots, e^{j \pi\left(N_{r,x}-1\right) \sin \theta \cos \phi}\right]^T
\end{equation}
\begin{equation}
\mathbf{u}_{el}(\phi)=\sqrt{\frac{1}{N_{r,y}}}\left[1, e^{j \pi \sin \phi}, \cdots, e^{j \pi\left(N_{r,y}-1\right) \sin \phi}\right]^T
\end{equation}
where $N_{t,x}$, $N_{t,y}$ and $N_{r,x}$, $N_{r,y}$ denote the number of transmit and receive antennas in each row and column of the UPA respectively. The beamforming vector at the $n$th slot $\mathbf{f}_{n}$ is designed based on the predicted angle $\hat{\bm{\theta}}_{n | n-1}$ from the $\left(n-1\right)$th slot as
\begin{equation}
\mathbf{f}_{n}=\mathbf{a}(\hat{\bm{\theta}}_{n | n-1})
\end{equation}

\subsection{Radar Measurement Model}
After sampling and performing block-wise Fourier Transform, the received discrete signal at the $m$th subcarrier, the $l$th symbol and the $i$th antenna can be expressed as:
\begin{equation}
\begin{aligned}
& r_{m,l} = \zeta \sqrt{p}\sum_{k=1}^K  \beta_{k} [\mathbf{b}(\bm{\theta}_{k,n})]_i \mathbf{a}^H(\bm{\theta}_{k,n}) \mathbf{f}_n s_{m,l}\\ 
& \times e^{j 2 \pi \mu_{k,n} \left(l-1\right) T_{s}} e^{-j 2 \pi \left(m-1\right) \Delta f \tau_{k,n} }
\end{aligned}
\end{equation}
where $i=0,...,N_r-1$. Noise is neglected for the sake of simplicity. The information of time delay and Doppler frequency need to be obtained accurately in order to estimate the range and velocity of targets. This can be tackled by performing an element-wise division between the received signal and the transmitted signal:
\begin{equation}
\begin{aligned}
& \Tilde{r}_{m,l} = \frac{r_{m,l}}{s_{m,l}}=\zeta \sqrt{p}\sum_{k=1}^K  \beta_{k} [\mathbf{b}(\bm{\theta}_{k,n})]_i \mathbf{a}^H(\bm{\theta}_{k,n}) \mathbf{f}_n\\ 
& \times e^{j 2 \pi \mu_{k,n} \left(l-1\right) T_{s}} e^{-j 2 \pi \left(m-1\right) \Delta f \tau_{k,n} }
\end{aligned}
\end{equation}

 By aggregating $M$ subcarriers and $L$ symbols, the processed signal at the $i$th antenna can be written into a compact matrix form as
 \begin{equation}
\begin{aligned}
\mathbf{\Tilde{R}}_{i} = \sum_{k=1}^K  \alpha_{k} \odot \bm{\eta}_{\tau_{k,n}} \bm{\omega}_{\mu_{k,n}}^H
\end{aligned}
\end{equation}
where $(\mathbf{\Tilde{R}}_{i})_{m,l}=\Tilde{r}_{m,l}$, $\alpha_{k}=\zeta \sqrt{p}\beta_{k} [\mathbf{b}(\bm{\theta}_{k,n})]_i \mathbf{a}^H(\bm{\theta}_{k,n}) \mathbf{f}_n$, and
\begin{equation}
\begin{aligned}
\bm{\eta}_{\tau_{k,n}} = \left[1, e^{-j 2 \pi \Delta f \tau_{k,n}}, \cdot, e^{-j 2 \pi \Delta f \left(M-1\right) \tau_{k,n}}\right]^T
\end{aligned}
\end{equation}
\begin{equation}
\begin{aligned}
\bm{\omega}_{\mu_{k,n}} = \left[1, e^{-j 2 \pi \mu_{k,n} T_{s}}, \cdot, e^{-j 2 \pi \mu_{k,n} \left(L-1\right) T_{s}}\right]^T
\end{aligned}
\end{equation}
 Thus, the acquisition of the range and velocity can be done by implementing 2D-DFT with respect to $\mathbf{\Tilde{R}}_{i}$. To be specific, by performing IFFT and FFT on the fast-time domain and slow-time domain respectively, the time delay and Doppler frequency can be estimated by detecting a peak at the corresponding index $(\hat{m}_k,\hat{l}_k)$ for each scatterer. The resulting distance and radial velocity can be given as
\begin{equation}
\begin{aligned}
\hat{d}_{k}=\frac{\hat{l}_k c}{2 M \Delta f}
\end{aligned}
\end{equation}
\begin{equation}
\begin{aligned}
\hat{v}_{k}=\frac{\hat{m}_k c}{2 f_c L T_{s}}
\end{aligned}
\end{equation}
where $f_c$ and $c$ denote the carrier frequency and the speed of light respectively.

To estimate the DOA, the MUSIC algorithm can be applied to the processed range profile at the $l$th symbol:
\begin{equation}
\begin{aligned}
\mathbf{\Tilde{D}}_{l}=[\mathbf{d}_1, ...,\mathbf{d}_{N_r-1}] \in \mathbb{C}^{M \times N_r}
\end{aligned}
\end{equation}
The covariance matrix can be written as $\mathbf{Y}=\frac{1}{L}\sum_{l=1}^L\mathbf{\Tilde{D}}_{l}^\mathbf{H}\mathbf{\Tilde{D}}_{l}$. After eigenvalue decomposition, it can be further expressed as:
\begin{equation}
\begin{aligned}
\mathbf{Y}=\mathbf{U}_s \bm{\Lambda}_s \mathbf{U}_s^H+\mathbf{U}_n \bm{\Lambda}_n \mathbf{U}_n^H
\end{aligned}
\end{equation}
where the diagonal matrices $\bm{\Lambda}_s$ and $\bm{\Lambda}_n$ each contains $K$ and $N_r-K$ eigenvalues, and the signal and noise subspace $\mathbf{U}_s$ and $\mathbf{U}_n$ contains $K$ and $N_r-K$ eigenvectors. The MUSIC spectrum can be given by 
\begin{equation}
\begin{aligned}
P_{MUSIC}(\bm{\theta})=\frac{1}{\mathbf{b}^{H}(\bm{\theta}) \mathbf{U}_n \mathbf{U}_n^H \mathbf{b}(\bm{\theta})} 
\end{aligned}
\end{equation}
By traversing all the possible directions of the receiving steering vector, the angle of the target can be estimated. Although the reflection coefficient is not measured in a direct manner, it can be calculated based on the measurement of $d_{k,n}$.

\subsection{Communication Model}
The received beamformed signal at the vehicle side can be formulated as
\begin{equation}
\begin{aligned}
c_n(t)=  \Tilde{\zeta} \sqrt{p}\sum_{k=1}^K  \Tilde{\alpha}_{k, n} \mathbf{v}_n \mathbf{u}(\bm{\theta}_{k, n}) \mathbf{a}^H(\bm{\theta}_{k, n}) \mathbf{f}_{n} \\ 
\times \sum_{m=0}^{M-1}\sum_{l=0}^{L-1} s_{m,l} e^{j 2 \pi m \Delta f \left(t-lT_s\right)} +z_c(t)
\end{aligned}
\end{equation}
where $\Tilde{\zeta} = \sqrt{N_tM_r}$ denotes the array gain factor, with $M_r$ being the number of receive antennas and $\Tilde{\alpha}_{k, n}$ denotes the channel coefficient of different paths. $\mathbf{u}\left(\bm{\theta}_{k, n}\right)$ is the receive steering vector and has the similar expression as (\ref{2}). The receive beamforming vector at the user side is derived based on the two step prediction as shown in Fig.\ref{fig2}, which is expressed as
\begin{equation}
\mathbf{v}_{n}=\mathbf{u}(\hat{\bm{\theta}}_{n | n-2})
\end{equation}
Here, we define the transmit signal-to-noise ratio (SNR) as
\begin{equation}
\begin{aligned}
\operatorname{SNR}_t= \frac{p}{{\sigma_r}^2} 
\end{aligned}
\end{equation}
and the receive SNR as
\begin{equation}
\begin{aligned}
\operatorname{SNR}_r= \frac{p\left|\Tilde{\zeta}\sum_{k=1}^K  \Tilde{\alpha}_{k, n} \mathbf{v}_n \mathbf{u}(\bm{\theta}_{k, n}) \mathbf{a}^H(\bm{\theta}_{k, n}) \mathbf{f}_{n}\right|^2}{{\sigma_r}^2} 
\end{aligned}
\end{equation}
where ${\sigma_r}^2$ denote the variances of the receive white Gaussian noise.

\section{5G NR Sensing-assisted communications} 

\subsection{NR Frame Structure}
Similar to LTE, 5G NR still uses OFDM waveform with cyclic prefix (CP). Different numerologies $\mu$ are used in NR to address different cases of frame structure with different subcarrier spacing\cite{3gpp.38.211}, which can be expressed as $\Delta f=2^\mu \cdot 15$ kHz, $\mu\in\mathbb{N}$, $\mu\leq6$. Although NR supports different numerologies, the lengths of one radio frame and one subframe are the same, which are 10ms and 1ms. Each subframe can be further divided into $2^\mu$ slots and each slot occupies 14 symbols for normal CP or 12 symbols for extended CP respectively. NR Dynamic Time Division Duplex (TDD) supports multiple slot formats that allow different symbols in one slot to be used for Downlink (DL), Uplink (UL) or Flexible, which provides more flexibility in different transmission situations.

\subsection{NR Beam Management}
Beam management is an essential part in NR link establishment and maintenance between User Equipments (UEs) and mmWave Next Generation Node Base (gNB), which is widely applied in initial access for idle users and beam tracking for connected users to provide optimal beam pairs, high beamforming gain and improve the throughput and quality of the communication\cite{3gpp.38.802}. 

{\it {1) Initial Access:}} Three stages can be summarized in initial access. During the first and second stages, gNB sweeps beamformed SS blocks and finer Channel State Information Reference Signal (CSI-RS) beams and finds the best beam based on the feedback of UE. Then gNB transmits the best finer beam repeatedly and UE performs beam sweeping to find the best receiving-end beam by measuring the power of the received CSI-RS. 

{\it {2) Connected Mode:}} After initial access, a data transmission link is formed between UE and gNB, thus the UE changes from idle mode to connected mode\cite{3gpp.38.804}. To help the UE obtain the transmission resource scheduling information, Downlink Control Information (DCI) is carried in the Physical Downlink Control Channel (PDCCH). The transmission data is mainly carried by Physical Downlink Shared Channel (PDSCH), in which different reference signals are mapped, such as DMRS, CSI-RS and Phase Tracking Reference Signal (PTRS).
\begin{itemize}
    \item {\it DMRS:} DMRS is mainly used by UE for coherent demodulation of PDSCH. Different mapping types, density and additional DMRS are supported. 
    \item {\it CSI-RS:} CSI-RS is utilized for downlink channel state information acquisition, whose configuration is flexible in NR. To be specific, up to 32 ports are supported and two types of codebook can be chosen from based on the number of ports, type of panels and number of users in MIMO. The report that sends back to gNB from UE contains parameters like rank indicator (RI), precoding matrix indicator (PMI), and channel quality information (CQI).
    \item {\it PTRS:} PTRS is introduced to compensate for the common phase error, which can be caused by the phase noise produced in local oscillators.
\end{itemize}

\begin{figure}[htbp]
\centering
\includegraphics[width=6cm]{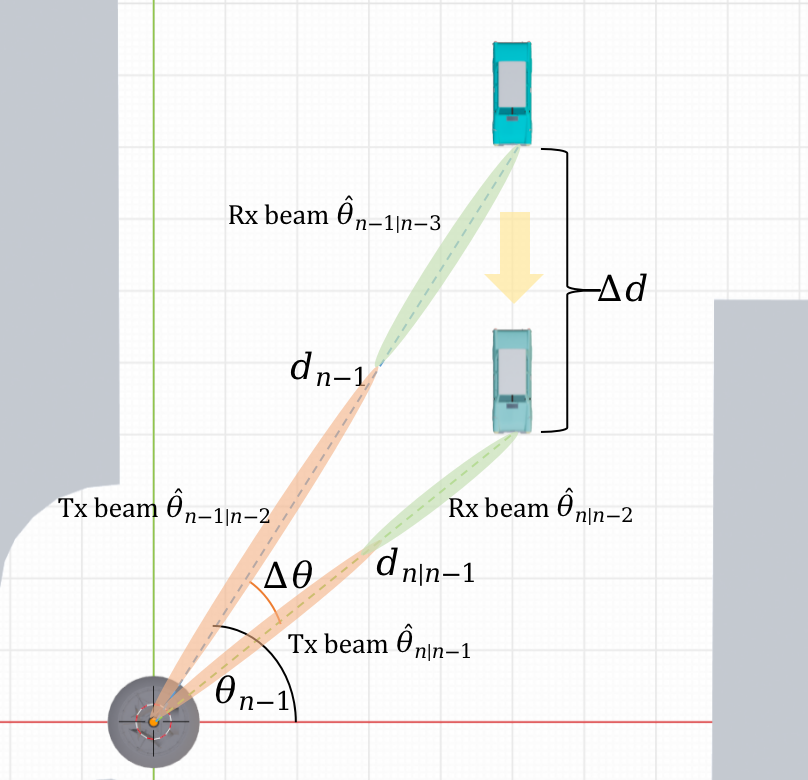}
\caption{State Evolution Model}
\label{fig2}
\end{figure}  

\subsection{ISAC in NR V2I network}
The above reference signals in NR connected mode are considered to be overheads since they occupy resource elements instead of data payloads. This may limit the effective communication performance, especially for high-mobility V2I links. Fortunately, with the utilization of ISAC in the NR V2I network, some of the reference signals can be reduced, thus improve the overall throughput. 

The CSI-RS mapped in PDSCH in connected mode is mainly used for channel estimation. To be specific, UE obtains the information of channel based on the received CSI-RS and sends feedback to gNB which contains the parameters that UE prefers, such as PMI and RI. The downlink CSI-RS and the uplink feedback are useful in designing the transmission scheme for the next period but they also cause considerable overheads which will reduce the throughput and data rate. However, this dilemma can be tackled by applying the ISAC signaling approach in the NR network, which provides the CSI information to the gNB based on the echo signals reflected by the vehicles. 

As an example, let us consider the scenario with a single layer SU-MIMO V2I network, where the feedback report of CSI-RS contains the information of PMI and CQI in conventional NR communication. However, by utilizing ISAC signal, gNB is capable of extracting essential motion information like motion parameters of the vehicle from the echo and analyze the channel quality based on the power of the echo, which can be further exploited in designing beamforming for the next epoch. Thus, CSI-RS can be abolished with the use of the ISAC NR signal in V2I networks, which reduces the overheads and improves the throughput during data transmission. As a step further, EKF can be exploited to predict and track the kinematic parameters of the vehicle. Following the derivation in \cite{liu2020radar} and based on the geometric relationships in Fig.\ref{fig2}, the state evolution model can be summarized as
\begin{equation}\label{16}
\left\{\begin{array}{l}
\theta_n=\theta_{n-1}-d_{n-1}^{-1} v_{n-1} \Delta T \cos \theta_{n-1}+\omega_\theta \\
d_n=d_{n-1}-v_{n-1} \Delta T \sin \theta_{n-1}+\omega_d \\
v_n=v_{n-1}+\omega_v \\
\beta_n=\beta_{n-1}\left(1-d_{n-1}^{-1} v_{n-1} \Delta T \sin \theta_{n-1}\right)^2+\omega_\beta
\end{array}\right.
\end{equation}
where $\theta_n$, $d_n$, $v_n$ and $\beta_n$ denote the angle, distance, velocity and reflection coefficient at the $n$th slot, respectively. The state evolution model and the measurement model can be formulated in compact forms as
\begin{equation}
\left\{\begin{array}{l}
\text {State Evolution Model: } \mathbf{x}_n=\mathbf{g}\left(\mathbf{x}_{n-1}\right)+\boldsymbol{\omega}_n \\
\text {Measurement Model: } \mathbf{y}_n=\mathbf{x}_n+\mathbf{z}_n
\end{array}\right.
\end{equation}
where $\mathbf{x}=[\theta, d, v, \beta]^T$ and $\mathbf{y}=[\hat{\theta}, \hat{d}, \hat{v}, \hat{\beta}]^T$ denote the state variable and the measurement variable respectively, $\mathbf{g}$ is defined in (\ref{16}), $\boldsymbol{\omega}=[\omega_\theta, \omega_d, \omega_v, \omega_\beta]^T$ and $\mathbf{z}=[z_\theta, z_d, z_v, z_\beta]^T$ are the zero-mean Gaussian noises caused by approximation and measurement respectively, whose covariance matrices can be expressed as 
\begin{equation}
\mathbf{Q}_s=\operatorname{diag}\left(\sigma_\theta^2, \sigma_d^2, \sigma_v^2, \sigma_\beta^2\right)
\end{equation}
\begin{equation}
\mathbf{Q}_m=\operatorname{diag}\left(\Tilde{\sigma}_\theta^2, \Tilde{\sigma}_d^2, \Tilde{\sigma}_v^2, \Tilde{\sigma}_\beta^2\right)
\end{equation}
The variances of the measurement noises are directly proportional to CRBs from \cite{liu2022survey}. To linearize the state evolution model, the Jacobian matrix of $\mathbf{g}$ can be given as:
\begin{small}
\begin{equation}
\begin{aligned}
\frac{\partial \mathbf{g}}{\partial \mathbf{x}}={\left[\begin{array}{cccc}
1+\frac{v \Delta T \sin \theta}{d} & \frac{v \Delta T \cos \theta}{d^2} & -\frac{\Delta T \cos \theta}{d} & 0 \vspace{0.8ex}\\
-v \Delta T \cos \theta & 1 & -\Delta T \sin \theta & 0 \vspace{0.8ex}\\
0 & 0 & 1 & 0 \vspace{0.8ex}\\
-\frac{2 \beta v \Delta T \cos \theta}{d}\iota & \frac{2 \beta v \Delta T \sin \theta}{d^2}\iota & -\frac{2 \beta \Delta T \sin \theta}{d}\iota & \iota^2
\end{array}\right]}
\end{aligned}
\end{equation}
\end{small}
where $\iota=\left(1-\frac{v \Delta T \sin \theta}{d}\right)$. Then the procedure of EKF can be summarized as follows.

{\it {1) State Prediction:}} 
\begin{equation}
\hat{\mathbf{x}}_{n \mid n-1}=\mathbf{g}\left(\hat{\mathbf{x}}_{n-1}\right), \hat{\mathbf{x}}_{n+1 \mid n-1}=\mathbf{g}\left(\hat{\mathbf{x}}_{n \mid n-1}\right).
\end{equation}

{\it {2) Linearization:}} 
\begin{equation}
\mathbf{G}_{n-1}=\left.\frac{\partial \mathbf{g}}{\partial \mathbf{x}}\right|_{\mathbf{x}=\hat{\mathbf{x}}_{n-1}}, \mathbf{H}_n=\mathbf{I}_4
\end{equation}

{\it {3) MSE Matrix Prediction:}} 
\begin{equation}
\mathbf{M}_{n \mid n-1}=\mathbf{G}_{n-1} \mathbf{M}_{n-1} \mathbf{G}_{n-1}^H+\mathbf{Q}_s
\end{equation}

{\it {4) Kalman Gain Calculation:}} 
\begin{equation}
\mathbf{K}_n=\mathbf{M}_{n \mid n-1} \mathbf{H}_n^H\left(\mathbf{Q}_m+\mathbf{H}_n \mathbf{M}_{n \mid n-1} \mathbf{H}_n^H\right)^{-1}
\end{equation}

{\it {5) State Tracking:}} 
\begin{equation}
\hat{\mathbf{x}}_n=\hat{\mathbf{x}}_{n \mid n-1}+\mathbf{K}_n\left(\mathbf{y}_n-\hat{\mathbf{x}}_{n \mid n-1}\right))
\end{equation}

{\it {6) MSE Matrix Update:}} 
\begin{equation}
\mathbf{M}_n=\left(\mathbf{I}-\mathbf{K}_n \mathbf{H}_n\right) \mathbf{M}_{n \mid n-1}
\end{equation}
where $\mathbf{I}_4$ denotes the identity matrix of size four. 

\subsection{Overhead Reduction Analysis}
\begin{figure}[htbp]
\centering
\includegraphics[width=\columnwidth]{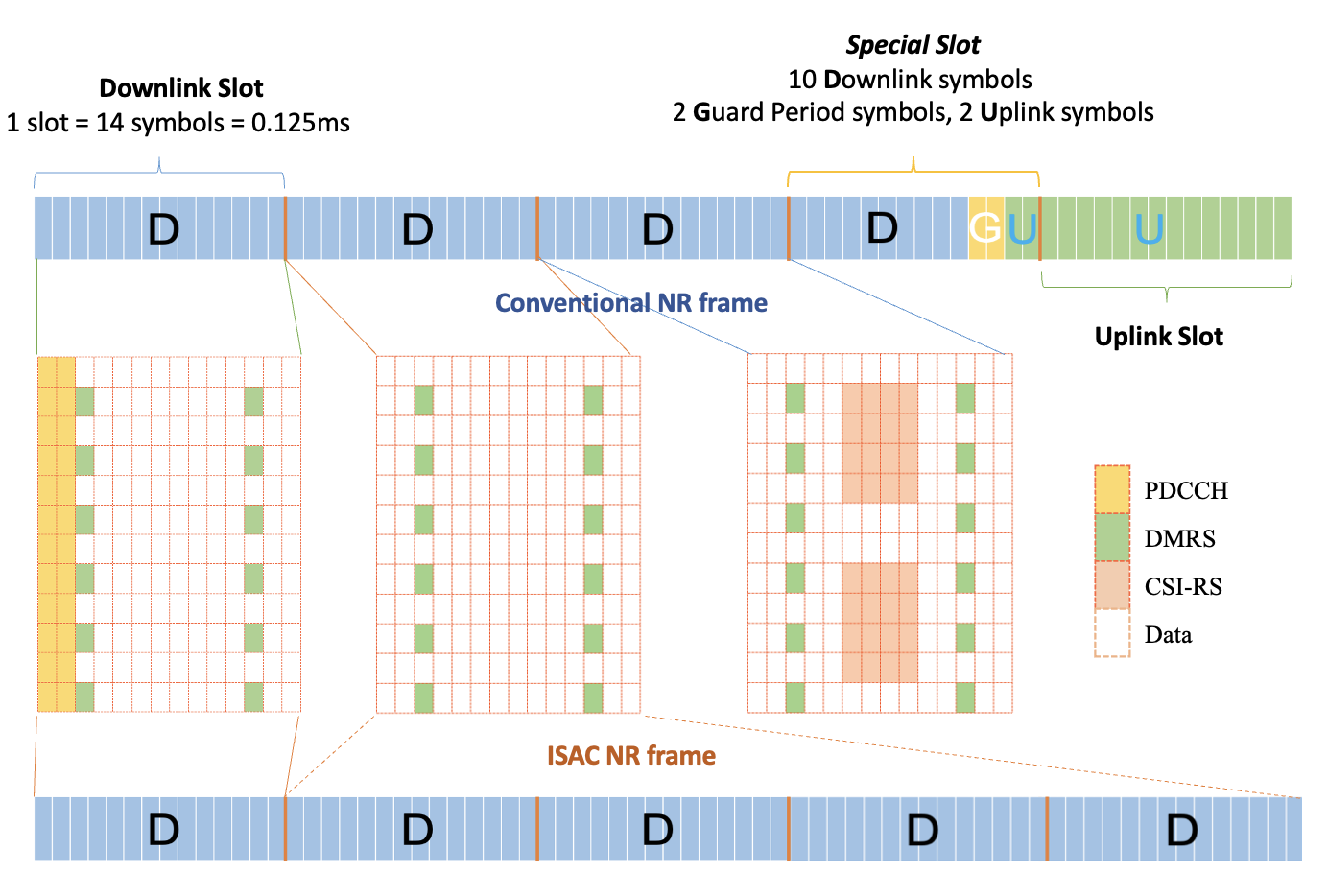}
\caption{Common Frame Structure}
\label{fig3}
\end{figure}

We assume that the V2I network adopts a common frame structure ``$DDDSU$'' in 5G NR with the numerology of $\mu=3$, where ``$D$'', ``$S$'' and ``$U$'' denote the downlink slot, special slot and uplink slot respectively. Also, we suppose that DMRS in PDSCH has the mapping type of ``$A$'' and we add an additional DMRS with the consideration of the high mobility of the target. In conventional NR frame, the CSI-RS occupies the maximum 32 antenna ports and is sent with the period of 5 slots. Thus, in ISAC NR frame structure, resource elements (RE) that CSI-RS occupies can be replaced by actual downlink data. The frame structures of conventional NR and ISAC NR in one period and one resource block are compared in Fig.\ref{fig3}. Based on the frame structure, the throughput of 5G NR can be expressed as:
\begin{equation}
\begin{aligned}
 \text{Throughput}\left(\text{in Mbps}\right) = 10^{-6}\cdot \sum_{j=1}^J\left(N_{\operatorname{Layers}}^{\left(j\right)} \cdot Q_m^{\left(j\right)} \right.\\
 \left. \cdot \frac{N_{PRB}^{BW\left(j\right),\mu}\cdot12}{T_s^\mu} \cdot \left(1-\operatorname{BER}^{\left(j\right)}-OH^{\left(j\right)}\right) \right)
\end{aligned}
\end{equation}
where $J$, $N_{Layers}$, $Q_m$, $N_{PRB}^{BW,\mu}$, $T_s^\mu$ denote the number of carriers in carrier aggregation,  number of layers in MIMO, modulation order, number of practical resource blocks, and the average OFDM symbol duration, respectively. Moreover, $\operatorname{BER}$ and $OH$ represent the bit error rate (BER) and the overhead percentage. 

From the conventional NR frame structure, we can see that the total downlink overheads can be considered as the resource elements (RE) that the reference signals occupy, which are 42 REs for DMRS and 32 REs for CSI-RS, respectively. By utilizing EKF to predict and track the UE, CSI-RS can be reduced in ISAC NR frame and the overheads can be reduced by $32/(42+32)=43.24\%$. In addition, the uplink feedback and the guard period are abolished under ISAC NR frame structure since ISAC NR V2I network is based on the analysis of the echo rather than the feedback. The resources that used to be occupied by guard period and uplink feedback in conventional NR communication can be thus used for downlink data transmission instead, which helps improve the downlink throughput and utilize spectral resources more efficiently.

\begin{table}[!h]
    \caption{Parameters of Simulation} \label{Sim} 
    \normalsize
    \centering
    \begin{tabular}{p{1.7cm} p{1.7cm} p{1.7cm} p{1.7cm}} 
    \hline
    \hline
    Parameter & Value & Parameter & Value\\
    \hline
    $f_c$ & $35$GHz & $T_{max}$ & $4\operatorname{s}$\\
    $\Delta f$ & $120$kHz & $\Delta T$ & $0.125\operatorname{ms}$\\
    $\sigma_\theta$ & $10^{-3}\operatorname{rad}$ & $\Tilde{\sigma}_\theta$ & $0.1\operatorname{rad}$\\
    $\sigma_d$ & $10^{-3}\operatorname{m}$ & $\Tilde{\sigma}_d$ & $0.2\operatorname{m}$\\
    $\sigma_v$ &$10^{-3}\operatorname{m/s}$ & $\Tilde{\sigma}_v$ & $0.15\operatorname{m/s}$\\
    $N_t, N_r$ & $64\;(8\times 8)$ & $M_r$ & $16\;(4\times 4)$\\
    $N_{\operatorname{Layers}}$ & $1$ & $Q_m$ & $4$\\
    $N_{PRB}^{BW,\mu}$ & $52$ & $T_s^\mu$ & $8.929\mu s$\\
    \hline
    \end{tabular}
\end{table}

\section{Simulation Results} 

The simulation scenario shown in Fig. \ref{fig1} is extracted from real building groups in Shenzhen. We assume that the vehicle is driving along a straight road across the building groups, and the wireless communication channel between the vehicle and the BS is a clustered delay line channel consisted of both LoS and NLoS paths. Both the transmit and receive physical antenna arrays are assumed to be UPA. In the simulated communication-only case, 32 CSI-RS antenna ports are adopted at BS and 4 subarrays are assumed in both horizontal and vertical directions in the transmit array, with 4 oversampled DFT beams in both directions. Suppose the BS is at the origin with the antenna height to be 4m and the vehicle starts at (25m, 40m) with the antenna height to be $1m$. Moreover, the initial velocity of the vehicle is 20m/s. Other simulation parameters are specified in Table. \ref{Sim}.

\begin{figure}[htbp]
\centering
\includegraphics[width=\columnwidth]{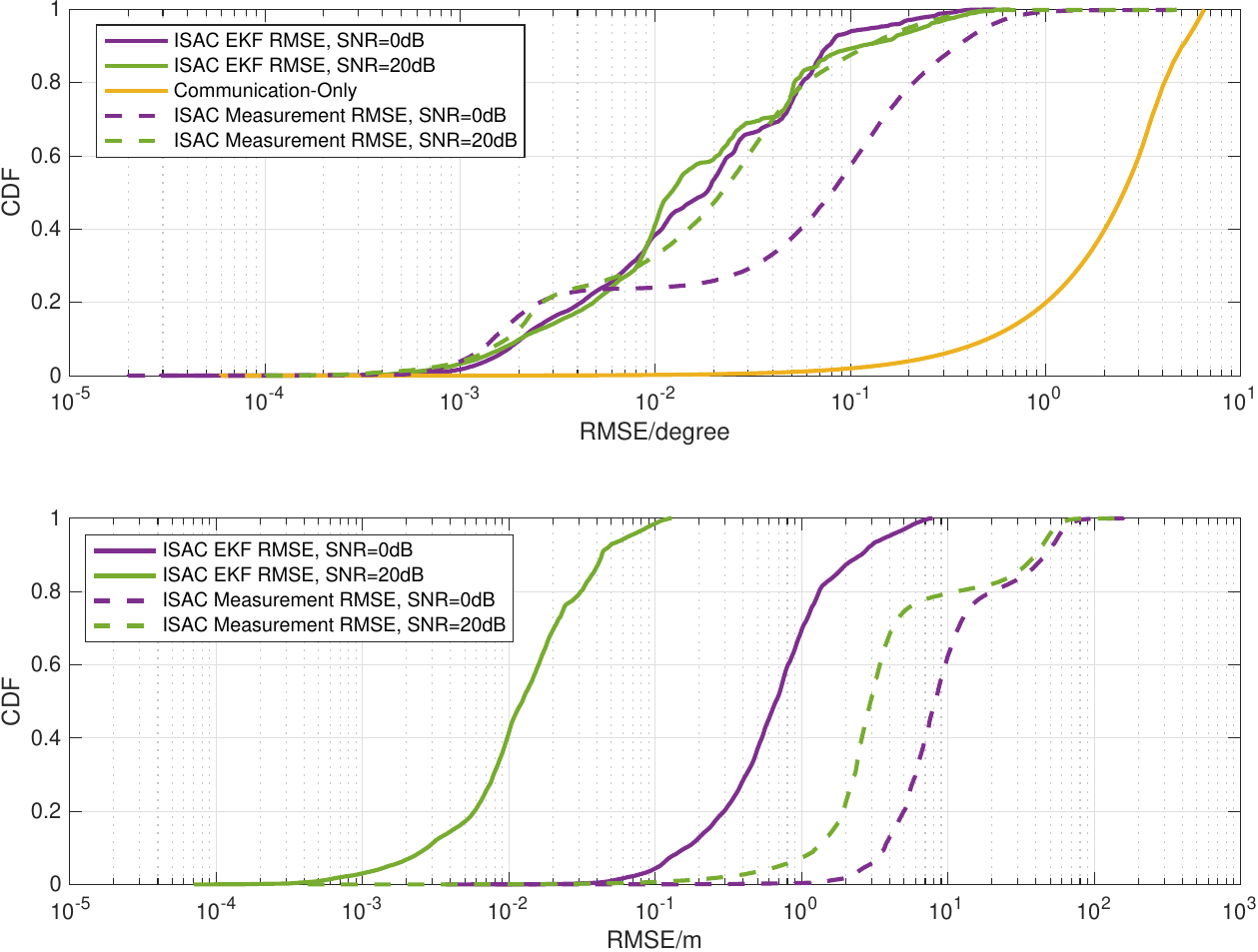}
\caption{CDF of RMSEs for different signals under different receive SNR}
\label{fig4}
\end{figure} 

In Fig. \ref{fig4}, we firstly explore the tracking performance for angle and distance by showing the empirical cumulative distribution functions (CDF) in terms of root mean squared error (RMSE) under different receive SNR. It can be seen that compared with the NR communication signal, the NR-based ISAC signaling achieves better tracking performance. That is because the communication-only signal follows the type-I CSI-RS codebook, which predefined a certain number of beamforming angles. Also, one can notice that after utilizing EKF in predicting and tracking, the RMSE is improved significantly compared to the RMSE that are obtained from the direct measurement of the echo, which shows the effectiveness of EKF in the considered V2I network.

\begin{figure}[htbp]
\centering
\includegraphics[width=\columnwidth]{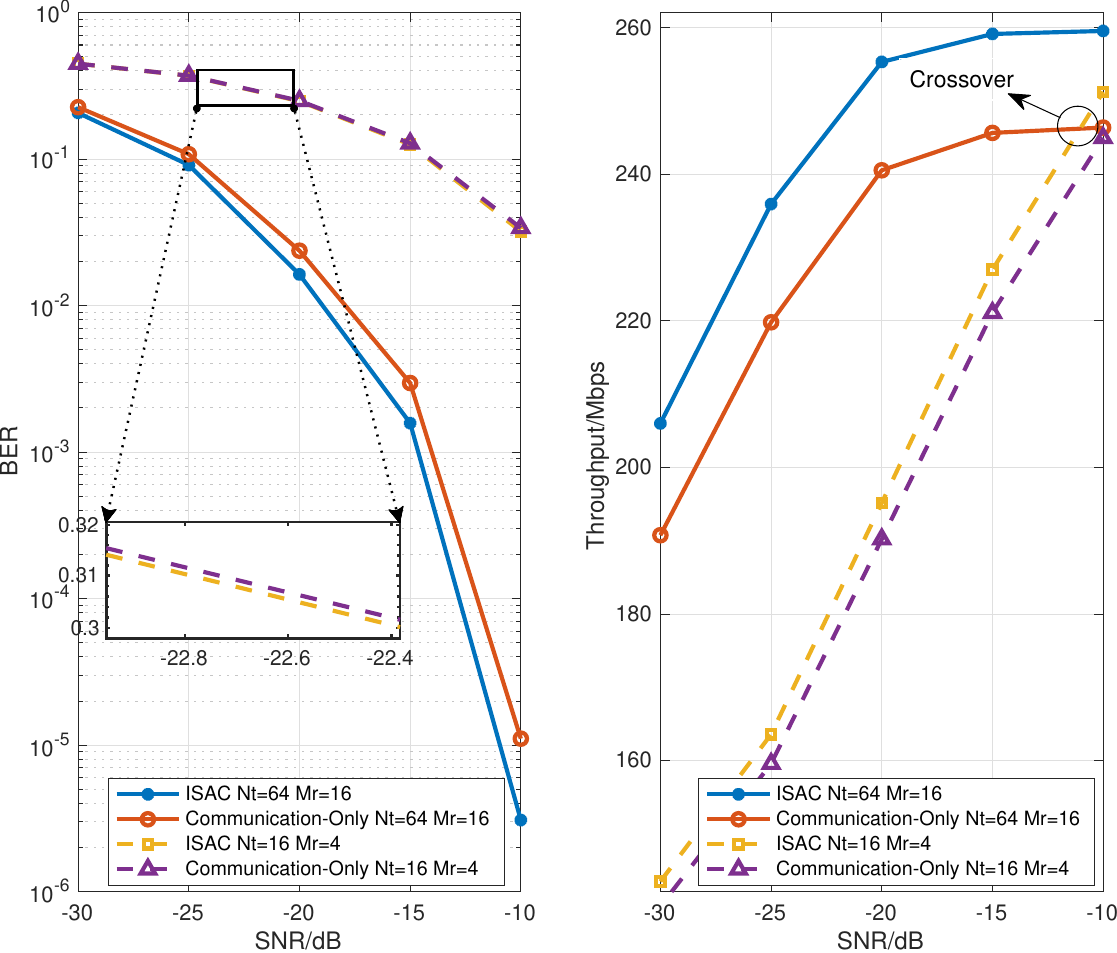}
\caption{BER and throughput under different transmit SNR for different signals}
\label{fig5}
\end{figure} 

In Fig. \ref{fig5}, we investigate the relationship between BER, throughput and transmit SNR under different number of transmit and receive antennas. With the increase of SNR, BER decreases and throughput gets improved. Compared with the NR communication-only signal, the NR-based ISAC signal shows slightly better performance under the evaluation of BER and clear improvement of throughput thanks to the reduction of the overheads. Moreover, significant performance boosts can be observed in both BER and throughput with the increase of antennas. Note that the crossover happens because only 16 CSI-RS ports are used in the case of $N_t=16$, which means the overheads are reduced compared to 32 CSI-RS ports in the case of $N_t=64$ and that leads to the higher throughput at high SNR.

\section{Conclusion} 
In this paper, we have proposed the utilization of NR ISAC signal in V2I network and predictive beam tracking approach to reduce the overheads caused by reference signal, e.g., CSI-RS. With the use of the sensing ability, EKF can be applied in beam tracking based on the measurement of the echo signal. The overhead reduction has been analyzed based on the frame structure of NR. Finally, numerical results have been provided to validate the improved performance of beam tracking and communication throughput.

\bibliographystyle{IEEEtran}
\bibliography{IEEEabrv,ref}
\end{document}